\documentclass[]{aa}     
\usepackage{amssymb}
\usepackage{amsmath}
\usepackage[mathcal]{eucal}
\usepackage{txfonts}
\usepackage{graphicx}

\usepackage[T1]{fontenc}

\usepackage{natbib}

\newcommand{\eqn} [1] {
\begin{equation}
#1
\end{equation}}

\newcommand{\be}{\begin{equation}}
\newcommand{\bea}{\begin{eqnarray}}
\newcommand{\ee}{\end{equation}}
\newcommand{\bc}{\begin{center}}
\newcommand{\ec}{\end{center}}
\newcommand{\bt}{\begin{table}}
\newcommand{\et}{\end{table}}
\newcommand{\eea}{\end{eqnarray}}
\newcommand{\bfig}{\begin{figure}}
\newcommand{\efig}{\end{figure}}
\newcommand{\bfige}{\begin{figure*}}
\newcommand{\efige}{\end{figure*}}

\def\gd {{$\gamma$~Doradus}}
\def\gds {{$\gamma$~Doradus~stars}}

\def\teff {{\mathrm{T}_{\mathrm{eff}}}}
\def\logg {{\log\,g}}

\def\vsini {{v\!\sin\!i}}

\def\muHz {{\mu\mbox{Hz}}}

\def\msol {{\mathrm{M}_\odot}}
\def\lsol {{\mathrm{L}_\odot}}

\def\hda {{HD\,12901}}

\def\vaiss {Brunt--Väisälä}

\def\iobs {${\cal I}_{\mathrm{obs}}$}
\def\iteo {${\cal I}_{\mathrm{th}}$}

\def\ppa {{Physical Process in Astrophysics}}

\begin{document}

   \title{Frequency ratio method for seismic modeling of $\gamma$ 
Doradus stars}

    \author{A. Moya\inst{1*} \and J.C. Su\'arez\inst{1,2} \and P. J. Amado 
\inst{1} \and S. Martin-Ru\'{\i}z\inst{1} \and R. Garrido\inst{1}}
    \institute{$^1$ Instituto de Astrof\'{\i}sica de Andaluc\'{\i}a (CSIC), Granada, 
               Spain\\
 	      $^2$ LESIA, Observatoire de Paris-Meudon, UMR8109, France}

    \date{Received ... / Accepted ...}
\offprints{A. Moya,\\
\email{moya@iaa.es}\\
$^*$ Current address: LESIA, Observatoire de Paris-Meudon, UMR8109,
France}

\abstract{A method for obtaining asteroseismological information of a
$\gamma$ Doradus oscillating star showing at least three pulsation
frequencies is presented. This method is based on a first-order
asymptotic g-mode expression, in agreement
with the internal structure of \gds. The information obtained is
twofold: 1) a possible identification of the radial order $n$ and
degree $\ell$ of observed frequencies (assuming that
these have the same $\ell$), and 2) an estimate of
the integral of the buoyancy frequency (Brunt-V\"ais\"al\"a) weighted
over the stellar radius along the radiative zone.  The accuracy of the
method as well as its theoretical consistency are also discussed for a
typical \gd\ stellar model. Finally, the frequency ratios method has
been tested with observed frequencies of the \gd\ star HD\,12901. The
number of representative models verifying the complete set of
constraints (the location in the HR diagram, the Brunt-V\"ais\"al\"a
frequency integral, the observed metallicity and frequencies and a
reliable identification of $n$ and $\ell$) is drastically reduced to
six.

}
\maketitle


\section{Introduction}

The asymptotic approximation for the solution of linear, isentropic
$g$-mode oscillations of high radial order was firstly attempted by
\cite{Sme68}, \cite{Tas68} and \cite{Zahn70}. These initial approaches
were developed for the special case of a star with a convective core
and a radiative envelope. From these first developments until
recently, all asymptotic solutions have been carried out by
using the \cite{Cow41} approximation, where the perturbation of the
gravitational potential is neglected.

\citet{Tas80} investigated the asymptotic representation of
high-frequency $p$-modes and low-frequency $g$-modes associated with
low-degree spherical harmonics. There, the procedure of \citet{Iwe68}
and \citet{Van67} was adopted to avoid the difficulties of the so-called 
mobile singularities.  This procedure solves different
equations in different regions of the star and gives asymptotic
expressions for the $p$ and $g$-modes for different stellar
structures. Later, \citet{Sme87} used pulsating second-order
differential equations, expressed in a single dependent variable, to make the 
construction of successive asymptotic approximations
more transparent.

\citet{Tas90} developed, for the first time, an asymptotic expression
for high-frequency non-radial $p$ modes without neglecting the
Eulerian perturbation of the gravitational potential. This study was
carried out for low-frequency $g$-modes by \citet{Sme95} by using a
different asymptotic approach. They applied a procedure described by
\citet{Kev81} to a fully radiative star, which was extended to a
convective core plus a radiative envelope by \citet{Wil97}.  Since the
work of \citet{Tas80}, subsequent studies have improved the second
order and the eigenfunction descriptions, but the first order
asymptotic expressions for the frequency in different physical
situations was correctly obtained by \citet{Tas80}.

The first attempt to obtain information on a real star by using
these $g$-mode asymptotic developments was carried out by \citet{Pro86}
adapting the second order asymptotic theory of \citet{Tas80} to the
special case of the Sun. They argued that the study of the $g$ modes
of a star would provide information about the physical conditions of
the stellar core.  At the same time, \citet{Kaw87} and \citet{Kaw94}
started to investigate the internal structure of the PG 1159 hot white dwarf
stars through the same asymptotic $g$-mode theory. By using the fully
radiative star expression, they obtained stellar properties comparing
mode differences of consecutive orders.

Recently, $\gamma$~Doradus stars have been defined by
\citet{Kaye99} as a class of stars pulsating in high-$n$, low-$\ell$ $g$
modes with very low photometric amplitude,
lying on or just to the right of the cold border at the lower part of
the Cepheid instability strip.  During recent years, tens of
candidate $\gamma$~Doradus stars have been observed and cataloged
\citep[][and references therein]{Math04,Sus02}.  Also,
several theoretical studies have been carried out to understand their
instability mechanism, since the $\kappa$-mechanism does not explain
the observed modes \citep[see][]{Guz00,ahmed,Ma04}.

Modelling these and other pulsating stars is mainly dependent on the
fundamental stellar parameters such as metallicity, effective
temperature, luminosity and surface gravity as deduced from
observations. In the case of the well-known five minute solar
oscillations, helioseismology has provided important improvements to
our knowledge of the internal structure and evolution of the Sun
thanks to the asymptotic theory applied to solar $p$ modes
\citep{jcd89}, useful for constraining the mass and the evolutionary stage of
solar-like pulsating stars by using the $(\Delta\sigma_0,D_0)$ diagrams. 
$\Delta\sigma_0$ represents the average
separation between modes of the same degree and adjacent radial order
and $D_0$ is related to the small separation between $\sigma_{n,\ell}$
and $\sigma_{n-1,\ell+2}$. This increases the set of
parameters ${\cal P}$ provided by classical observables with the
seismic information
\be
{\cal P}={\cal P}(Z, g, L, \teff; \Delta\sigma_0, D_0).\nonumber
\ee
where $Z$ represent the relative metal abundance, $g$ the gravity,
$L$ the luminosity and $\teff$ the effective temperature of
the star.
                       
An analogous method for low frequency asymptotic $g$-mode pulsators,
such as \gds, will be given here.  Additional
constraints can be obtained by using information inferred from ratios
of the observed frequencies.  Similarly to acoustic modes in the
asymptotic regime, our method increases the set of
parameters ${\cal P}$ provided by classical observables by
\be
{\cal P}={\cal P}(Z, g, L, \teff; \frac{f_{o,i}}{f_{o,j}}, {\cal I}).\nonumber
\ee
where $f_{o,i}$ and $f_{o,j}$ represent the observed frequencies and
${\cal I}$ is the integral of the Brunt-V\"ais\"al\"a frequency
defined below.  Particularly, this procedure will provide an estimate
of the radial and spherical orders $(n,\ell)$ of at least
three observed frequencies of a $\gamma$~Doradus star.  In addition,
the internal stellar structure will also be constrained through the
knowledge of the integral along the stellar radius of the
Brunt-V\"ais\"al\"a frequency.

The paper is organized as follows: In Sec. 2, the first order
analytical asymptotic equation to be used is presented, as well as the
physical reasons for the choice of this expression. In Sec. 3, the
procedure and a test of its theoretical consistency is discussed. An
application of the method to observations of the \gd\ star \hda\
is given in Sec. 4.  

\section{Analytical equation for asymptotic $g$ modes}

$\gamma$ Doradus stars are found to pulsate in the low frequency
g-mode asymptotic regime.  There are several analytical forms for
obtaining the numerical value of the pulsational period as a function
of the radial order $n$, the spherical order $\ell$, different
constants and equilibrium quantities, all of them having slight
differences and are applied to different internal stellar
structures.

\begin{figure}
	\resizebox{\hsize}{!}{\includegraphics{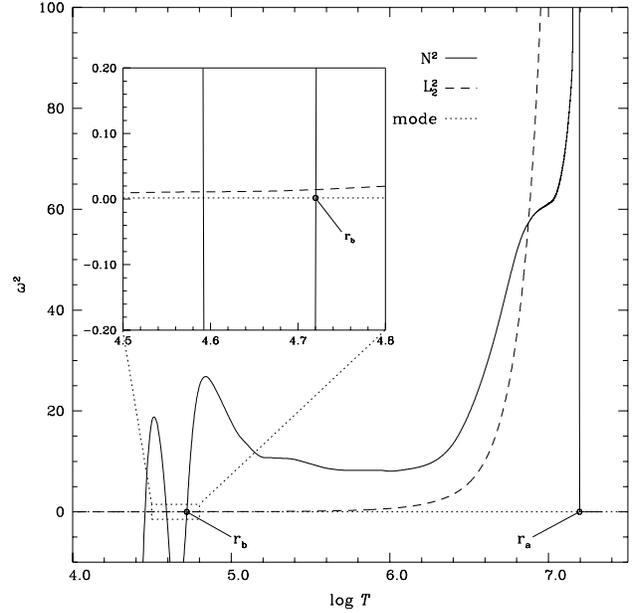}}

      \caption{Propagation diagram for the theoretical model presented
               in Table \ref{freq_mod}. Angular frequencies are given in
               nondimentional units. The horizontal line
               represent a standard $\gamma$ Doradus frequency, around
               $20\,\mu Hz$ ($\omega^2$= 0.0018).}
      \label{prop}
\end{figure}
These stars present a convective core, a radiative envelope and a
small outer convective zone close to the photosphere.  The
distribution of convective and radiative zones inside the star
produces different propagation cavities, which is the key for solving
the problem of interpreting the behavior of their $g$-modes in the
asymptotic regime.  The so-called \emph{turning points} represent the
limits of such cavities. They are defined, depending on the authors, by
the zeros of the \vaiss\ frequency \citep{Tas80} or by the points of
intersection of the propagating mode with the \vaiss\ and Lamb
frequencies, whichever comes first \citep{Shi79} (see
Fig. \ref{prop}). Both quantities represent the
characteristic buoyancy (\vaiss) and acoustic (Lamb) frequencies 
\citep{Unno89}.

For the \gd\ internal distribution of convective and radiative zones,
the better adapted analytical solution is
\be
\sigma^{a}_{n,\ell}\approx
\frac{[\ell(\ell+1)]^{1/2}}{(n+1/2)\pi}{\cal I}
\label{Tassoul}
\ee
given by \citet{Tas80} and obtained under the assumption of
adiabaticity and non-rotation. $\cal I$ is the integral of the
Brunt-V\"ais\"al\"a frequency, given by
\be
{\cal I}=\int_{r_a}^{r_b}\frac{N}{r}\mathrm{d}r~,
\label{eq:defI}
\ee
where $r_a$ and $r_b$ represent the inner and outer turning points
respectively.


%

%

\begin{figure}
	\resizebox{\hsize}{!}{\includegraphics{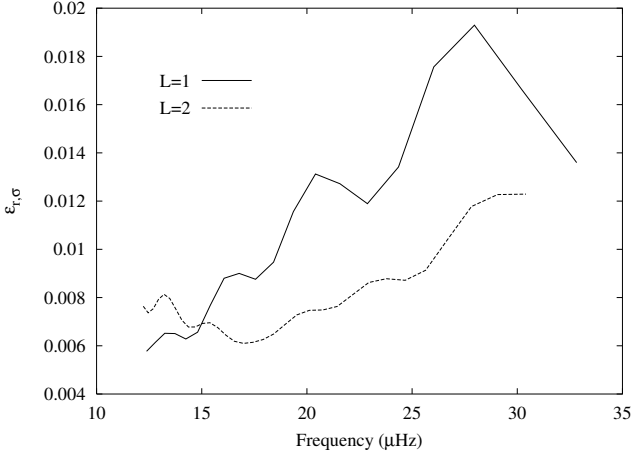}}

    \caption{Relative errors of asymptotic frequencies compared
             to theoretical predictions ($\epsilon_{r,\sigma}$) by
             using Eq.~\ref{Tassoul} for a complete theoretical
             oscillation spectrum. See text for details.}
\label{frec_the}
\end{figure}

Considering low $g$-mode frequencies in the asymptotic regime (typical 
of \gds), the inner turning point ($r_a$) is
 clearly defined by the zero of $N^2$
(Fig.~\ref{prop}).  However, for the outer layers,
small changes in the frequency produce different turning points
($r_b$) depending on whether the frequency intersects first $N^2$ or
$L^2_2$ (see Fig.~\ref{prop}) compared to the zero of $N^2$.
The problem of considering one of these possibilities
remains unsolved. This theoretical question beyond the scope of the
present work, however an estimate of the error committed when
considering Eq.~\ref{Tassoul} can be calculated.  To do so,
theoretical oscillation frequencies $\sigma_{n,\ell}$ computed from a
given model (see Table~\ref{freq_mod}) are compared with those
obtained from Eq.~\ref{Tassoul}. Such errors, defined as
\be
\epsilon_{r,\sigma}=\Big|\frac{\sigma_{n,\ell}-\sigma^a_{n,\ell}}
                          {\sigma_{n,\ell}}\Big|,\label{eq:defepsilonsig}
\ee
are shown in Fig.~\ref{frec_the} for $\ell\!\!=\!1$ and $\ell=2$ modes. As
expected, $\epsilon_{r,\sigma}$ increases with the mode frequency 
and hence decreases with $n$, since they depart from the asymptotic regime. 
In the range of frequencies in which \gds\ pulsate, these 
errors remain around 1--2\%.

Equivalently, the relative error committed when calculating ${\cal I}$
values from Eq.~\ref{Tassoul} can be investigated through
\be
\epsilon_{r,{\cal I}}=\Big|\frac{{\cal I}_{\mathrm{th}}-{\cal I}^a_i}
{{\cal I}_{\mathrm{th}}}\Big|,\label{eq:defepsilonI}
\ee
where ${\cal I}^a_i$ is calculated for each mode provided the
frequency and the radial and spherical orders of the modes, and \iteo\
represents the integral ${\cal I}$ calculated for the theoretical
model.  Errors obtained in this case are of the same order as
$\epsilon_{r,\sigma}$, and the average value of $1\%$ observed in
Fig. \ref{frec_the} will be used in Sec. 4.  They remain lower than
typical errors given by the resolution of observed frequencies and
hence, of their ratios.  This represents an important advantage for
the method since no significant additional uncertainties need to
be considered besides those coming from observations.

Therefore, considering all previous arguments, the analytic form
(Eq.~\ref{Tassoul}) given by \citet{Tas80} will be adopted in this
work.
%
\section{Frequency Ratio method\label{sec:themethod}}

\subsection{Method bases\label{ssec:fundmethod}}

In the case of asymptotic acoustic modes, physical information can be
inferred from frequency differences of adjacent radial order
modes. This can be done since the model dependence appears in the
equation as an adding term. As can be seen from the analytic form for
asymptotic $g$ modes (Eq.~\ref{Tassoul}), it is not possible to retrieve
physical information from frequency differences.  However, the role
played by such frequency differences can be replaced by frequency
ratios.

Let us consider two modes $\sigma_{\alpha_1}$ and $\sigma_{\alpha_2}$,
with $\alpha_1\equiv({n_1},{\ell_1})$ and $\alpha_2 \equiv
({n_2},{\ell_2})$ respectively. For sake the of simplicity, the same
mode degree $\ell_1=\ell_2=\ell$
of a non-rotating pulsating star, showing adiabatic $g$ modes in
asymptotic regime, is considered. Their eigenfrequencies can be
approximated by Eq.~\ref{Tassoul} which is model-dependent through the
integral given by Eq.~\ref{eq:defI}. As shown in Fig.~\ref{prop}, for
typical \gd\ frequency ranges, the possible slight differences of the outer
turning point location can be considered negligible as far as the
calculation of ${\cal I}$ is concerned. It can thus be approximated as
constant (${\cal I}_{\alpha_2}={\cal I}_{\alpha_1}={\cal I}$), and
therefore, the ratio of $\sigma_{\alpha_1}$ and $\sigma_{\alpha_2}$
can be approximated by
\be
\frac{\sigma_{\alpha_1}}{\sigma_{\alpha_2}}\approx
\frac{n_2+1/2}{n_1+1/2}\;\; .
\label{cociente}
\ee
An estimate of the radial order ratios of these frequencies can thus
be obtained. A value of the integral can also be deduced from
observations ${\cal I}_{obs}$, provided that an estimate of the mode
degree $\ell$ is assumed. Constraints on models will come from the
consistency of: 1) the radial order identification corresponding to
the observed ratios; 2) their corresponding order $\ell$ and 3) the
observed \vaiss\ integral.

On the other hand, the assumption of equal
$\ell$ for all the observed modes is not imperative for this
procedure. Additional information on $\ell$ provided by
spectroscopy or multicolor photometry can be exploited through
the following expression \citep[see][]{Tas80}
\be
\frac{\sigma_{{n_1},{\ell_1}}}{\sigma_{{n_2},{\ell_2}}}\approx
\frac{n_2+1/2}{n_1+1/2}\frac{\sqrt{\ell_1(\ell_1+1)}}{\sqrt{\ell_2(\ell_2+1)}}
\label{cociente_con_l}
\ee
which is the extended form of Eq.~\ref{cociente}.

The efficiency and utility of this method depend on the number of
observed frequencies N$_f$. The larger N$_f$, the lower the number
of integers verifying Eq.~\ref{cociente}. The method
becomes useful for N$_f\ge3$.

\subsection{Theoretical test}

To test the self-consistency of this technique,
consider a theoretical numerical simulation of a given star. Assuming
three theoretical frequencies computed for a $1.5\msol$ typical \gd\
model on the Main Sequence, the ratio method is applied to
obtain an estimate of the radial order $n$ and the spherical order
$\ell$ of these frequencies as well as their corresponding ${\cal I}$.
If this procedure is self-coherent, at least one of all possible
solutions will correspond to the selected theoretical frequencies and
the corresponding ${\cal I}$ value.
\bt
\caption{$\ell$, $n$ and frequency $\sigma$ in $\mu Hz$ for a complete
spectrum of the theoretical model displayed at the top of the table.}
\begin{tabular}{rrrrrr} \hline\hline
\noalign{\medskip}
$\log\,\teff$ & $\log\,g$ & $\log\frac{L}{L_{\odot}}$ &$X_c$&
$\frac{M}{M_{\odot}}$&\iteo\\ 
   \noalign{\smallskip}
\hline 
   \noalign{\smallskip}
3.845 & 4.045 & 0.873 & 0.29 & 1.5 & 839\\
   \noalign{\smallskip}
\hline
   \noalign{\smallskip}
$\ell$ & $n$ & $\sigma$ $(\mu Hz)$ & $\ell$ & $n$ & $\sigma$ $(\mu Hz)$\\
   \noalign{\smallskip}
\hline
   \noalign{\smallskip}
1 & 30 & 12.3052 & 1 & 22 & 16.6262\\
1 & 29 & 12.7176 & 1 & 21 & 17.4038\\
1 & 28 & 13.1589 & 1 & 20 & 18.2397\\
1 & 27 & 13.6375 & 1 & 19 & 19.1344\\
1 & 26 & 14.1553 & 1 & 18 & 20.1369\\
1 & 25 & 14.7063 & 1 & 17 & 21.2964\\
1 & 24 & 15.2893 & 1 & 16 & 22.6059\\
1 & 23 & 15.9219 & 1 & 15 & 24.0274\\
   \noalign{\smallskip}
\hline
   \noalign{\smallskip}
2 & 53 & 12.1277 & 2 & 40 & 16.0399\\
2 & 52 & 12.3621 & 2 & 39 & 16.4502\\
2 & 51 & 12.5999 & 2 & 38 & 16.8789\\
2 & 50 & 12.8444 & 2 & 37 & 17.3283\\
2 & 49 & 13.1012 & 2 & 36 & 17.8010\\
2 & 48 & 13.3737 & 2 & 35 & 18.2982\\
2 & 47 & 13.6612 & 2 & 34 & 18.8212\\
2 & 46 & 13.9616 & 2 & 33 & 19.3752\\
2 & 45 & 14.2724 & 2 & 32 & 19.9675\\
2 & 44 & 14.5931 & 2 & 31 & 20.6010\\
2 & 43 & 14.9264 & 2 & 30 & 21.2734\\
2 & 42 & 15.2772 & 2 & 29 & 21.9839\\
2 & 41 & 15.6484 & 2 & 28 & 22.7434\\
   \noalign{\smallskip}
\hline
   \noalign{\smallskip}
\end{tabular}
\label{freq_mod}
\et
The theoretical eigenfrequencies (computed by the code FILOU, see
Sec.~4.2) of a typical \gd\ model, as well as its ${\cal I}_{th}$ value
in $\mu Hz$ are given in Table~\ref{freq_mod}.

\begin{figure}
	\resizebox{\hsize}{!}{\includegraphics{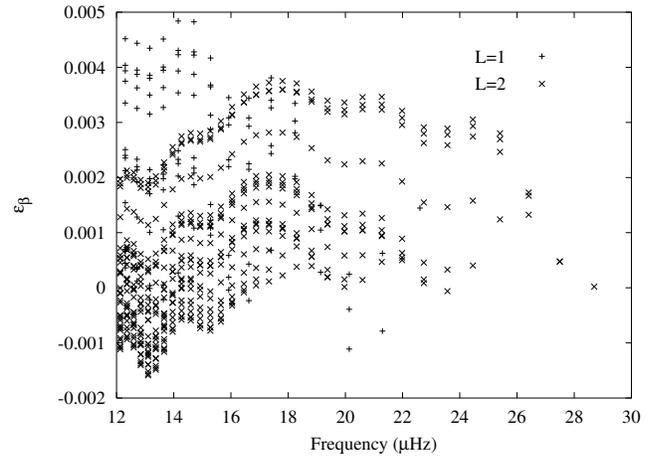}}
\caption{Error $\epsilon_{\beta}$ between the frequency ratios and the
corresponding asymptotic predictions, for the complete spectrum of the
theoretical model.}
\label{accu}
\end{figure}

A first test of the ratio method is carried out for three close
$\ell=1$ modes, for instance $(n,\sigma)= (25,14.706)$, $(23,15.922)$
and $(20,18.240)$. The corresponding frequency ratios are
\be
\frac{\sigma_1}{\sigma_2}=0.9236,
~~\frac{\sigma_2}{\sigma_3}=0.8729,
~~\frac{\sigma_1}{\sigma_3}=0.8063
\label{cocientes_teor}
\ee
which constitute our assumed \emph{observed} frequency ratios.  As in
this theoretical test the radial order of selected frequencies is
known, it is possible to study the accuracy of Eq.~\ref{cociente} by
comparing the corresponding $n$ ratios:
\be
\frac{n_2+1/2}{n_1+1/2}=\beta_1=0.9216,~~\beta_2=0.8723,~~ \beta_3=0.8039
\label{betas}
\ee
yielding an error of $2\times 10^{-3}$, $6\times 10^{-4}$ and $24\times
10^{-4}$ respectively.
In Fig.~\ref{accu}, the accuracy of Eq.~\ref{cociente}, defined as
\be
\epsilon_{\beta}=\frac{\sigma_i}{\sigma_j}-\frac{n_j+1/2}{n_i+1/2}
\label{eq:def_epsbeta}
\ee
is shown for all the frequencies of Table~\ref{freq_mod}.  The maximum
departure of the $n$ ratio from the real frequency ratios is $5\cdot
10^{-3}$, allowing us to consider this value as the error when
applying this method.
%
\subsubsection{Radial order identification}

If the values of the radial orders are assumed unknown and their
errors are supposed to be given by the maximum of $\epsilon_{\beta}$ 
(Eq.~\ref{eq:def_epsbeta}), i.e. $5\cdot 10^{-3}$, ratios of
integer numbers are searched for within the following ranges
\[\frac{\sigma_1}{\sigma_2}=[0.929,0.919]\]
\begin{equation}
\frac{\sigma_2}{\sigma_3}=[0.878,0.868]
\label{cocientes_teor2}
\end{equation}
\[\frac{\sigma_1}{\sigma_3}=[0.811,0.801]\]
Integer numbers contemplated here (from 1 to 60) cover the range of
typical $g$-mode radial orders shown by \gds. Within this range, there
exist several possible integer numbers yielding the same value for
each ratio of Eq.~\ref{cocientes_teor2}. However, the key point
consists of searching for ($n_1,n_2,n_3$) sets verifying
\be
\frac{n_2+0.5}{n_1+0.5}\sim\frac{\sigma_1}{\sigma_2},
~~\frac{n_3+0.5}{n_2+0.5}\sim\frac{\sigma_2}{\sigma_3},~~
\frac{n_3+0.5}{n_1+0.5}\sim\frac{\sigma_1}{\sigma_3}
\label{cocientes_teor3}
\ee
and limiting thus the final number of valid sets.

%
\bt
\caption{Radial number $n$ for each selected frequency, $\ell$ and
Brunt-V\"ais\"al\"a integral (in $\mu$Hz) estimated by the frequency ratio
method for the assumed {\em observed} frequencies $(20,18.2397)$,
$(23,15.9219)$ and $(25,14.7063)$ with $\ell=1$. The real values of the
radial and spherical orders are displayed in boldface.}
\begin{tabular}{rrrrrr} \hline\hline
\noalign{\medskip}
$n_1$ & $n_2$ & $n_3$ & $\ell$ &\iteo & $\epsilon_{\beta}$\\ 
   \noalign{\smallskip}
\hline 
   \noalign{\smallskip}
{\bf 20} & {\bf 23} & {\bf 25} & {\bf 1} & 831 & 0.0017\\
33 & 38 & 41 & 1 & 1360 & -0.0007\\
33 & 38 & 41 & 2 & 784 & -0.0007\\
41 & 47 & 51 & 2 & 971 & 0.0003\\
21 & 24 & 26 & 1 & 702 & -0.0035\\
42 & 48 & 52 & 2 & 801 & -0.0023\\
34 & 39 & 42 & 1 & 1127 & -0.0039\\
34 & 39 & 42 & 2 & 651 &  -0.0039\\
40 & 46 & 50 & 2 & 764 & 0.0030\\
   \noalign{\smallskip}
\hline 
\end{tabular}
\label{coc_1_cerc}
\et
Once all possible sets have been obtained, an estimated spherical
order $\ell$ to which $n_1$, $n_2$ and $n_3$ are associated can also
be identified.  To do so, selected sets are compared with theoretical
oscillation spectra computed from representative models of \gds.
Finally, for each solution ($\sigma,n,\ell$) found, an estimate of
\iteo\ (${\cal I}$ is approximately constant within each set) can also
be deduced through Eq.~\ref{Tassoul}.

In this exercise seven possible sets have been obtained. The real $n$
and $\ell$ (our simulated observations) have been retrieved (boldface
in Table~\ref{coc_1_cerc}) together with other eight possible
identifications. In addition, an accurate prediction of the
Brunt-V\"ais\"al\"a frequency is also obtained. To
investigate other distributions of observed frequencies (always within
the asymptotic regime), the following cases are studied: 1) modes with
very different radial order, for instance $(n_1=16,n_2=23,n_3=28)$ and
$\ell=1$; 2) close consecutive radial orders, for instance
$(n_1=45,n_2=47,n_3=50)$ and $\ell=2$; and 3) very different radial
orders $(n_1=30,n_2=45,n_3=53)$ and $\ell=2$.  Following the same
procedure, sets of 6, 5 and 8 possible solutions are obtained. For all
of them, the real values of $n$ and $\ell$ (boldface in
Table~\ref{coc_1}), and an estimate of ${\cal I}$ very close to the
values from the model are obtained.

These results are very important since they show the self-consistency
of the method which, in turn, reduces drastically the number of
representative models of a given \gd\ star. Moreover, they illustrate
the good accuracy of Eq.~\ref{Tassoul} for this kind
of study.

\bt
\caption{Radial number $n$ for each selected frequency, $\ell$ and
Brunt-V\"ais\"al\"a integral estimated by the frequency ratio method
for the assumed {\em observed} frequencies $(16,22.6059)$,
$(23,15.9219)$ and $(28,13.1589)$ with $\ell=1$ (set 1), $(45,14.2724)$,
$(47,13.6612)$ and $(50,12.8444)$ with $\ell=2$ (set 2) and $(30,21.2734)$,
$(45,14.2724)$ and $(53,12.1277)$ with $\ell=2$ (set 3), See text for details.
 The real radial and spherical orders are displayed in boldface.}
\begin{tabular}{rrrrrrr} \hline\hline
\noalign{\medskip}
set & $n_1$ & $n_2$ & $n_3$ & $\ell$ &\iteo & $\epsilon_{\beta}$\\ 
   \noalign{\smallskip}
\hline 
   \noalign{\smallskip}
1 & 26 & 37 & 45 & 2 & 768 & -0.0001\\
 & 33 & 47 & 57 & 2 & 971 & -0.0004\\
 & {\bf 16} & {\bf 23} & {\bf 28} & {\bf 1} & 830 & 0.0024\\
 & 23 & 33 & 40 & 1 & 1182 & 0.0013\\
 & 23 & 33 & 40 & 2 & 682 & 0.0013\\
 & 30 & 43 & 52 & 2 & 886 & 0.0007\\
   \noalign{\smallskip}
\hline
\noalign{\medskip}
set & $n_1$ & $n_2$ & $n_3$ & $\ell$ &\iteo & $\epsilon_{\beta}$\\ 
   \noalign{\smallskip}
\hline 
   \noalign{\smallskip}
2 & 46 & 48 & 51 & 2 & 850 & -0.0020\\
 & {\bf 45} & {\bf 47} & {\bf 50} & {\bf 2} & 832 & -0.0007\\
 & 43 & 45 & 48 & 2 & 798 & 0.0021\\
 & 44 & 46 & 49 & 2 & 815 & 0.0006\\
 & 47 & 49 & 52 & 2 & 869 & -0.0033\\
\noalign{\medskip}
\hline 
   \noalign{\smallskip}
set & $n_1$ & $n_2$ & $n_3$ & $\ell$ &\iteo & $\epsilon_{\beta}$\\ 
   \noalign{\smallskip}
\hline 
   \noalign{\smallskip}
3 & 22 & 33 & 39 & 1 & 1063 & 0.0005\\
 & 22 & 33 & 39 & 2 & 614 & 0.0005\\
 & 26 & 39 & 46 & 1 & 1252 & 0.0001\\
 & 26 & 39 & 46 & 2 & 723 & 0.0001\\
 & {\bf 30} & {\bf 45} & {\bf 53} & {\bf 2} & 833 & -0.0004\\
 & 34 & 51 & 60 & 2 & 941 & -0.0002\\
 & 33 & 49 & 58 & 2 & 910 & -0.0016\\
 & 18 & 27 & 32 & 1 & 876 & 0.0008\\
   \noalign{\smallskip}
\hline 
\end{tabular}
\label{coc_1}
\et
%

\section{An application: The \gd\ star \hda\label{sec:hd12901}}

Once the \emph{frequency ratio} method has been shown to be
self-consistent, a test with real data must be carried out for the
\gd\ star \hda. This star pulsates with three frequencies 
(Table~\ref{tab:obsdata}) in the asymptotic regime 
(see Sec.~\ref{sec:themethod} for more details).
\bt \caption{Photometric observed frequencies in both cycles per day
    and $\muHz$ for the \gd\ star \hda, taken from
    \citet{AertsCuypers04}.}  \vspace{1em}
    \renewcommand{\arraystretch}{1.2} \begin{tabular}[h]{cccc}
    \hline\hline & $\nu$ & $\nu$ \\ & ($c/d$) & ($\muHz$) \\ \hline &
    & \\ $f_{\rm I}$ & 1.216 & 14.069 \\ $f_{\rm II}$ & 1.396 & 16.157
    \\ $f_{\rm III}$ & 2.186 & 25.305 \\ \hline \end{tabular}
    \label{tab:obsdata} \et
%

\subsection{Fundamental parameters\label{ssec:fundparam}}

The fundamental parameters of \hda\ have been determined by
applying \emph{TempLogG} \citep{TemplogG} to the Str\"omgren--Crawford
photometry listed in the Hauck-Mermilliod catalogue
\citep{Hauc98}. In this catalogue, no $\beta$ index value is given
for \hda, which was instead obtained from \citet{Hand99}.  The code
classifies this object as a main sequence star in the spectral 
region F0-G2.  The resulting physical parameters are summarized in
Table~\ref{tab:params}.

\bt
     \caption{Physical parameters of the stars \hda taken
     from the literature.  The values of the rotational velocities are
     also listed.}
     \label{tab:params}    
  
     \vspace{1em}
     \renewcommand{\arraystretch}{1.2}
     \begin{tabular}{cccrcl}
     \hline\hline
HD & $\teff$ & $\logg$ & \multicolumn{1}{c}{${\rm
     [Fe/H]}$} & $\vsini$ &
     \multicolumn{1}{l}{Ref.}\\
    & (K) & (dex) & \multicolumn{1}{c}{(dex)} & (km\,s$^{-1}$) &\\
     \hline
12901 & {\bf 6996} & {\bf 4.04} & $-0.37$ &    & $a$\\
       & 7079 & 4.47 & $-0.40$ &    & $b$\\
       &      &      &         & 53 & $c$\\
       &      &      &         & 66 & $d$\\       
     \hline
     \end{tabular}
\footnotesize       

$a$: \emph{TempLogG} \citep{TemplogG}\\
$b$: \citet{Dupr02}\\
$c$: \citet{AertsCuypers04}\\
$d$: \citet{Math04}\\
  
\et
\citet{Dupr02} gives also physical parameters for \hda\ based on the
Geneva photometry of \citet{AertsCuypers04} and the calibrations given
by \citet{Kunzli97} for the Geneva photometry of B to G stars. In his
study, theoretical stellar models for this star do not take into
account their surface gravity determinations, as they are too high,
corresponding to models below the ZAMS.

Values for the $\vsini$ were taken from \citet{Roye02}, computed from
spectra collected at the Haute-Provence Observatory (OHP) and by
\citet{AertsCuypers04}, derived from a cross-correlation function
analysis of spectroscopic measurements with the CORALIE
spectrograph. \citet{Math04} have published the results
of a two-year high-resolution spectroscopic campaign, monitoring
59 \gd\ candidates. In this campaign, more than 60\% of the stars
presented line profile variations which can be interpreted as due to
pulsation.  From this work, in which projected rotational velocities
were derived for all 59 candidates, an additional value of $\vsini$ is
obtained for \hda.

\bfige
  \bc
\includegraphics[width=15cm]{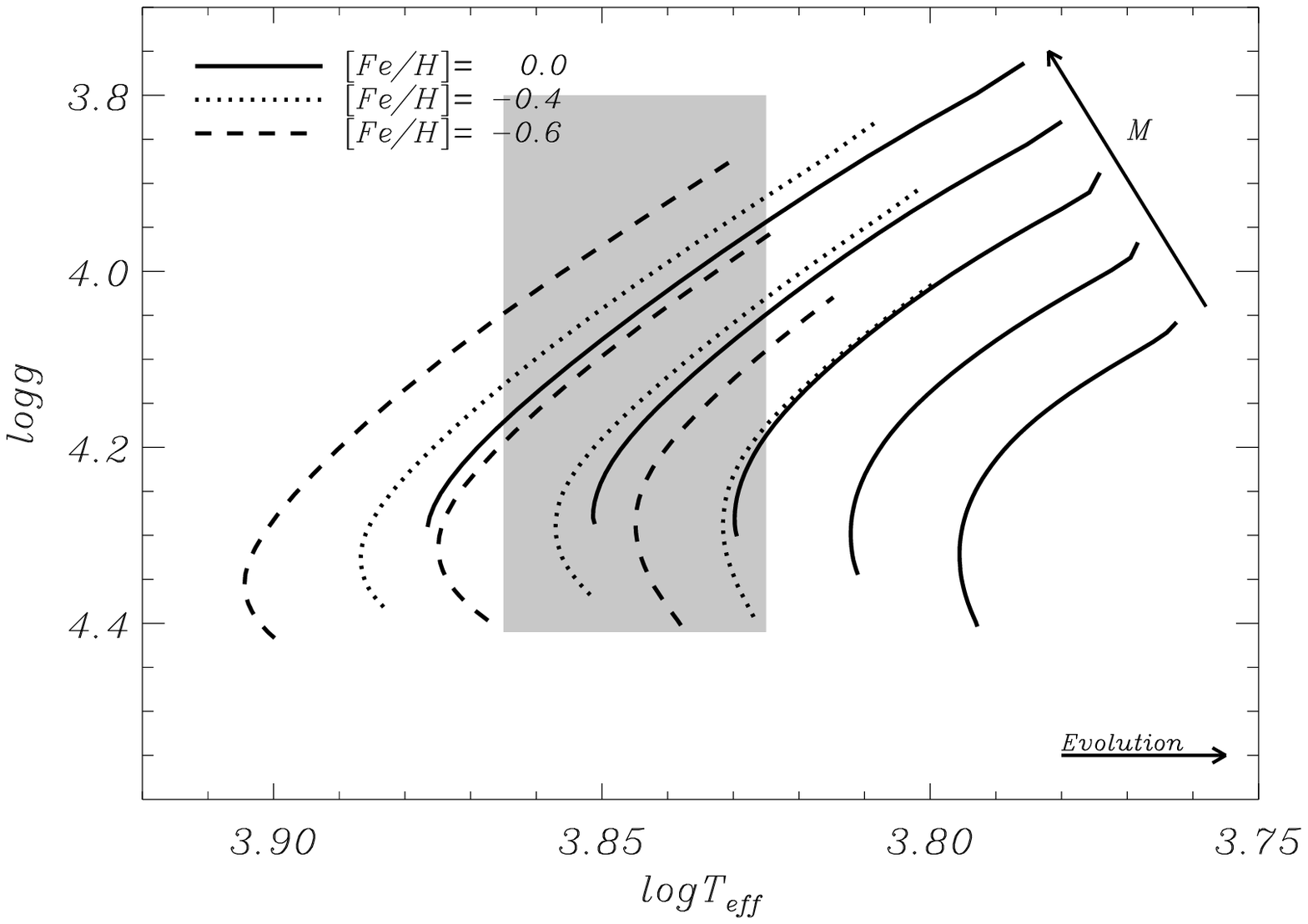}
   
   \caption{Evolutionary tracks of models representative of \hda\ in a
            $g$--$\teff$ (in a logarithmic scale). The shaded area 
	    represents the observational error box as deduced from
	    Table~\ref{tab:params}. From bottom to top, masses of models
	    vary from 1.2 to $1.6\msol$. Continuous lines represent
	    evolutionary tracks of models computed with [Fe/H]=0.0
	    (the solar value). Dotted lines represent those computed
	    with [Fe/H]=-0.4 and dashed lines are those 
	    computed with [Fe/H]=-0.6.}
	    \label{fig:logg-teff_general}
   \ec
\efige
The physical parameters used in the present work to delimit the error
boxes for the star in the HR diagram are those given by
\emph{TempLogG}, derived from the Str\"omgren--Crawford photometry and
given in blold font in Table~\ref{tab:params}.  Errors for these
parameters will be assumed to be approximately $0.2$~dex in
$\logg$ and $200$~K in $\teff$.

\subsection{Modelling\label{ssec:models}}

The evolutionary code CESAM \citep{Morel97} has been used to compute
stellar equilibrium models. The physics have been chosen as adequate
for intermediate mass stars. Particularly, the opacity tables are
taken from the OPAL package \citep{Igle96}, complemented at low
temperatures ($T\leq10^3\,K$) by the tables provided by
\citet{AlexFergu94}.  The convective transport is described by the
classical ML theory, in which the free parameters
$\alpha_{ML}=l_{m}/\mathrm{H}p=1.8$ and a mixed core overshooting
parameter $d_{ov}=l_{ov}/\mathrm{H}p=0.2$ are considered \citep[as
prescribed by][for intermediate mass stars]{Scha92}.  The
$\mathrm{H}p$ corresponds to the local pressure scale-height.  $l_{m}$
and $d_{ov}$ represent the mixing length and the inertial penetration
distance of convective bulbs respectively.  For the atmosphere
reconstruction, Eddington's $T(\tau)$ law (grey approximation) is
used. The transformation from heavy element abundances
with respect to hydrogen [M/H] into concentration in mass $Z$ assumes
an enrichment ratio of $\Delta Y/\Delta Z=2$ and
$Y_{\mathrm{pr}}=0.235$ and $Z_{\mathrm{pr}}=0$ as helium and heavy
element primordial concentrations.

To cover observational errors in the determination of
$\teff$, $\logg$ and [Fe/H], equilibrium models are computed within a
mass range of 1.2--$1.6\msol$ and metallicities from -0.6 up to the
solar value (Fig.~\ref{fig:logg-teff_general}).  Theoretical
oscillation spectra are computed with the oscillation code FILOU
\citep[see][]{filou,SuaThesis}.  Eigenfrequencies are computed for
mode degree values of up to $\ell=2$, thus covering the range of
observed $g$ modes for \gds\ (up to $g_{60}$).

\subsection{Locating \hda\ in a ${\cal I}$ -- $\log \teff$
            diagram \label{ssec:I-teff}}

\subsubsection{Observed frequency ratios\label{sssec:obs_ratios}}

Following the steps described in Sec.~\ref{sec:themethod}, the ratios of observed
frequencies are:
\eqn{\frac{f_{\rm I}}{f_{\rm II}}=0.871,~~
     \frac{f_{\rm II}}{f_{\rm III}}=0.639,~~
     \frac{f_{\rm I}}{f_{\rm III}}=0.556.\label{eq:freqratobs}}
On the other hand, all possible integer number ratios up to $n=60$ are
calculated. An error of $5\cdot10^{-3}$ on ratios values is assumed
in order to match the theoretical uncertainties found in
Fig.~\ref{accu}. With these assumptions and considering only dipoles
($\ell =1$) and quadrupoles ($\ell =2$), as has usually been found
in this type of star, all possible sets ($\sim60$) were reduced to
six, given in Table~\ref{tab:n-I-obs}. For each set, the
spherical order $\ell$ is estimated by comparing their radial orders
with theoretical spectra for representative models (see
Sec.~\ref{ssec:models} and Table. 1). Furthermore, it is possible to
obtain the \emph{observed} \vaiss\ integral (${\cal
I}_{obs}$) through Eq.~\ref{Tassoul}, constituting one of the main
constraints of this method.
\bt
  \caption{List of selected integer numbers sets associated to the 
           observed frequency ratios of \hda. For each set, the 
	   corresponding \emph{observed} ${\cal I}_{obs}$ is given in $\muHz$.
	   An estimate of the spherical order $\ell$ is
	   given in column 4. }
  \begin{tabular}{cccccr} \hline\hline
  \noalign{\medskip}
& $n_1$ & $n_2$ & $n_3$ & $\ell$ & ${\cal I}_{\mathrm{obs}}$\\ 
   \noalign{\smallskip}
\hline 
   \noalign{\smallskip}
$t_{1}$& 17 &  27 &  31 & 1 & 987.0\\
$t_{2}$& 21 &  33 &  38 & 1 & 1202.4\\
$t_{3}$& 21 &  33 &  38 & 2 & 694.2\\
$t_{4}$& 26 &  41 &  47 & 2 & 860.0\\
$t_{5}$& 30 &  47 &  54 & 2 & 984.3\\
$t_{6}$& 33 &  52 &  60 & 2 & 1087.9\\
 \hline
 \end{tabular}
  \label{tab:n-I-obs}
\et
\bfige
  \bc
   \includegraphics[width=15cm]{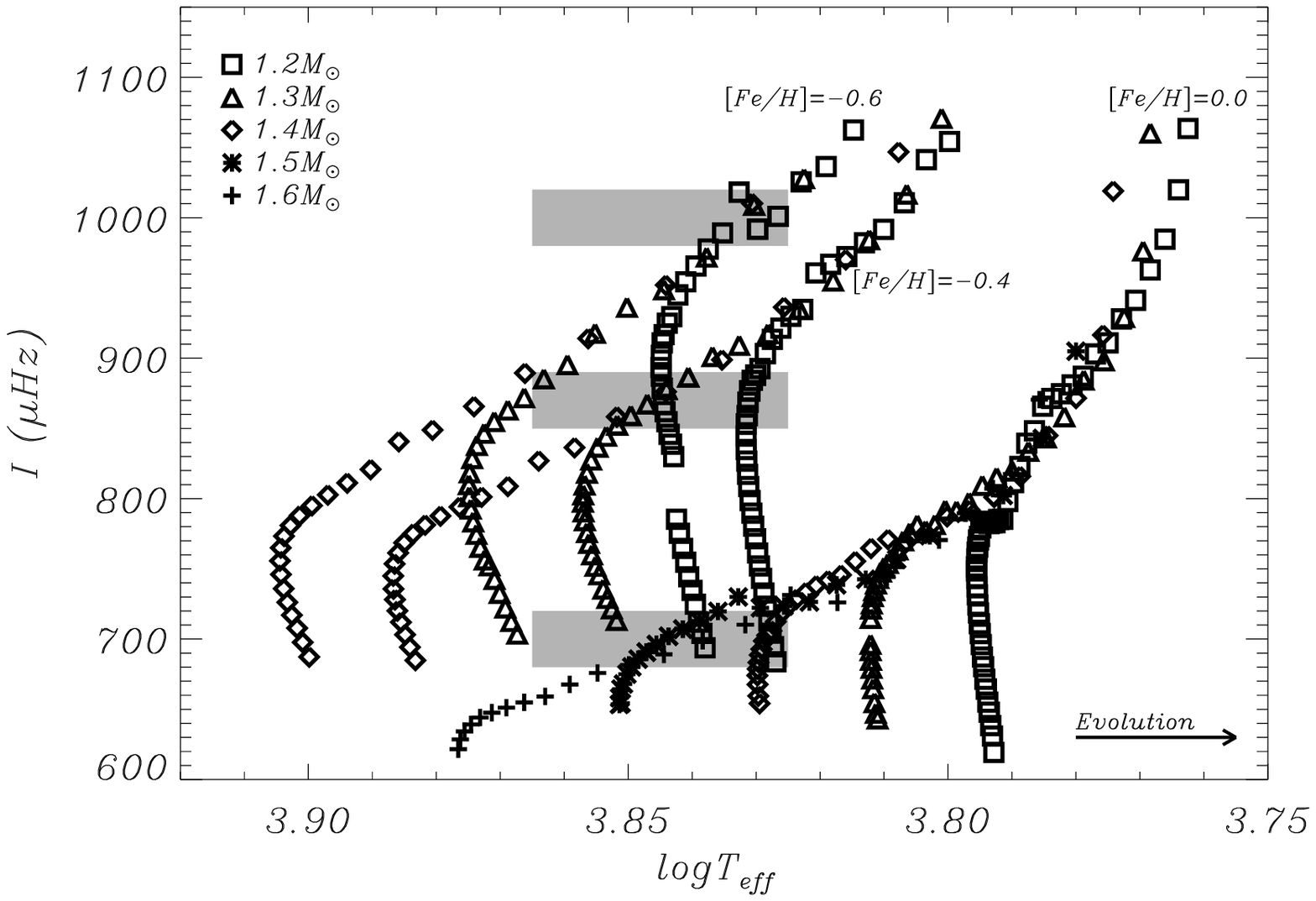}
   
   \caption{$\cal I$ as a function of the effective temperature for 
            representative models of \hda with masses in the range of 
	    $M=1.2$--1.6$\msol$, for three different metallicities
	    $[Fe/H]=$0.0, -0.4 and -0.6. Shaded areas represent uncertainty
	    (${\cal I}_{\mathrm{obs}}$,$\log \teff$) boxes of selected sets 
	    $t_1$, $t_3$ and $t_4$ as given in Table~\ref{tab:n-I-obs}.}
   \label{fig:int-teff}	    
   \ec
\efige
%
\subsubsection{The ${\cal I}$--$\log \teff$ diagram \label{sssec:I-teff_diag}}

For each valid set of integer numbers ($t_i$ in
Table~\ref{tab:n-I-obs}) there is only one associated ${\cal
I}_{obs}$ value, as deduced from Eq.~\ref{Tassoul}. Selected \iobs\
can thus be located on a theoretical ${\cal I}$--$\log \teff$
diagram.  This is done in Fig.~\ref{fig:int-teff} for the selected
$t_i$.  Only $t_1$, $t_3$ and $t_4$ are displayed since $t_2$ and
$t_6$ give a \vaiss\ integral value impossible to reproduce with the
theoretical models (see Fig.~\ref{fig:int-teff}), and $t_5$ has a
${\cal I}_{obs}$ impossible to differentiate from $t_1$ in the ${\cal
I}$--$\log \teff$ diagram. The theoretical ${\cal I}$ is computed for
a set of representative models of \hda\
(Fig.~\ref{fig:logg-teff_general}) in the range of 1.2--$1.6\msol$ for
three different metallicities ([Fe/H]= 0.0, $-0.4$ and $-0.6$).

Models are computed within an error box that takes into consideration
the typical accuracy for the calculation of physical parameters.  As
discussed in Sec. 2, a numerical uncertainty in ${\cal I}$ of 1\% is
chosen.

\bfige
   \includegraphics[width=9cm]{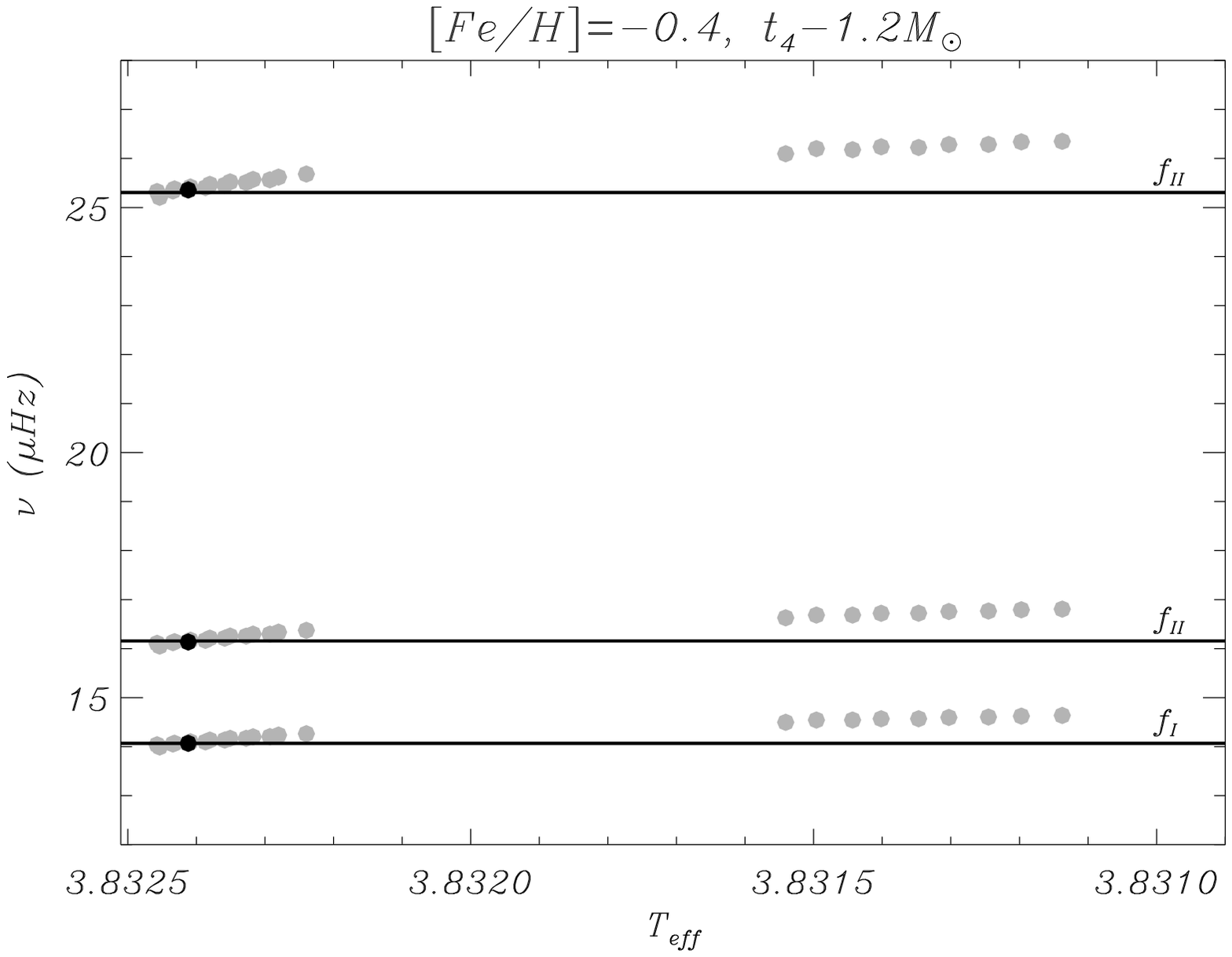}\hspace{-0.5cm}
   \includegraphics[width=9cm]{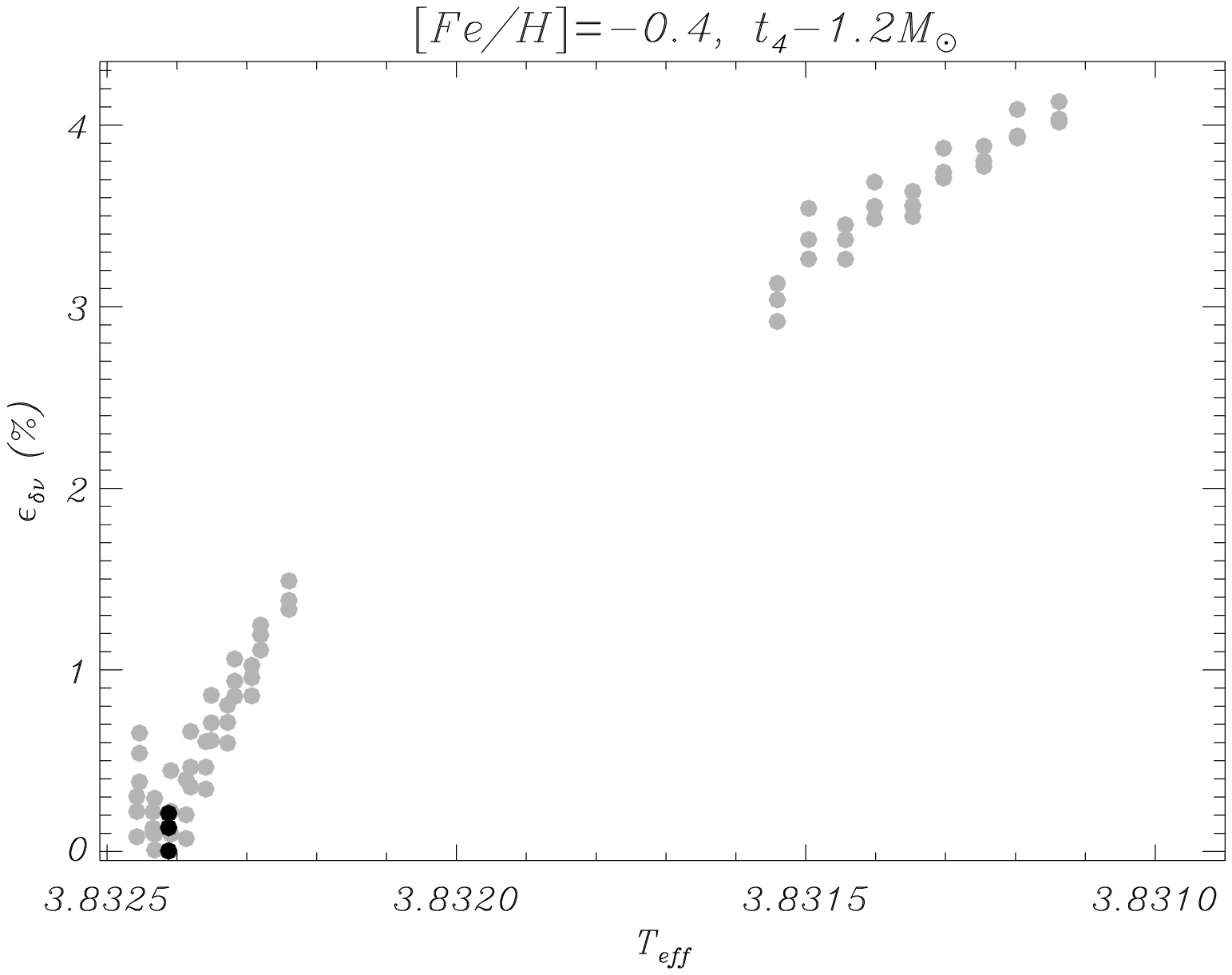}\hspace{-0.5cm}
   \includegraphics[width=9cm]{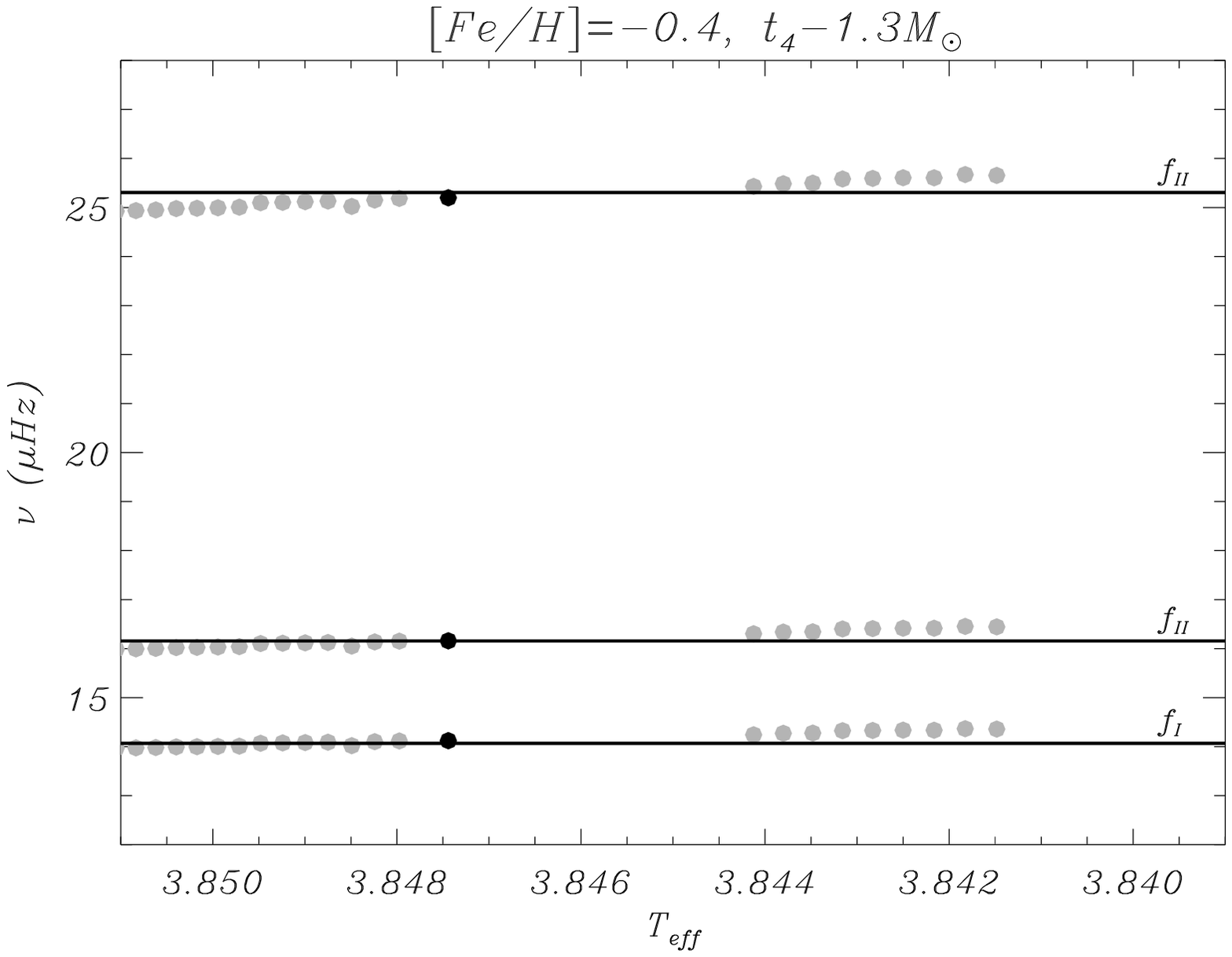}\hspace{-0.5cm}
   \includegraphics[width=9cm]{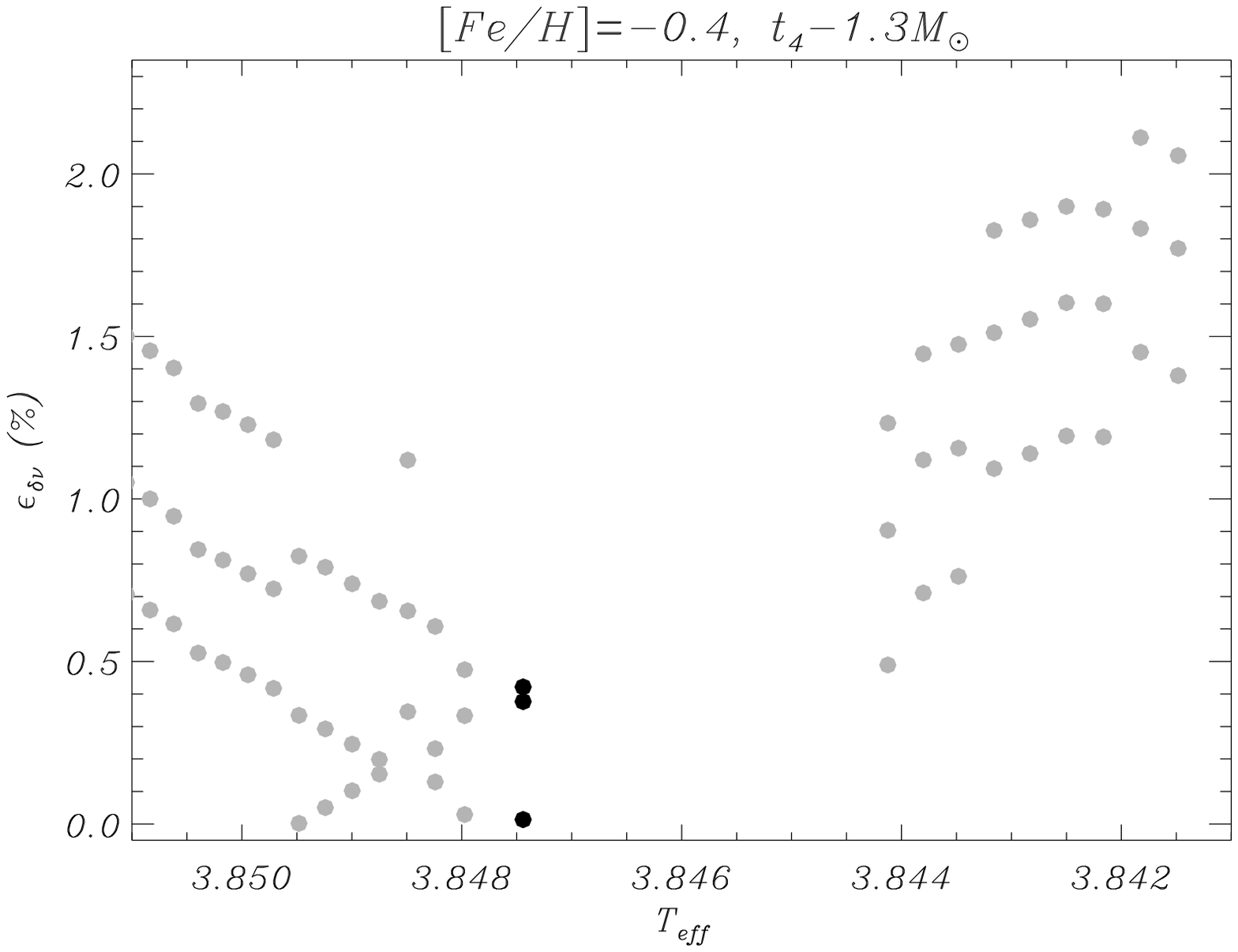}\hspace{-0.5cm}
   \includegraphics[width=9cm]{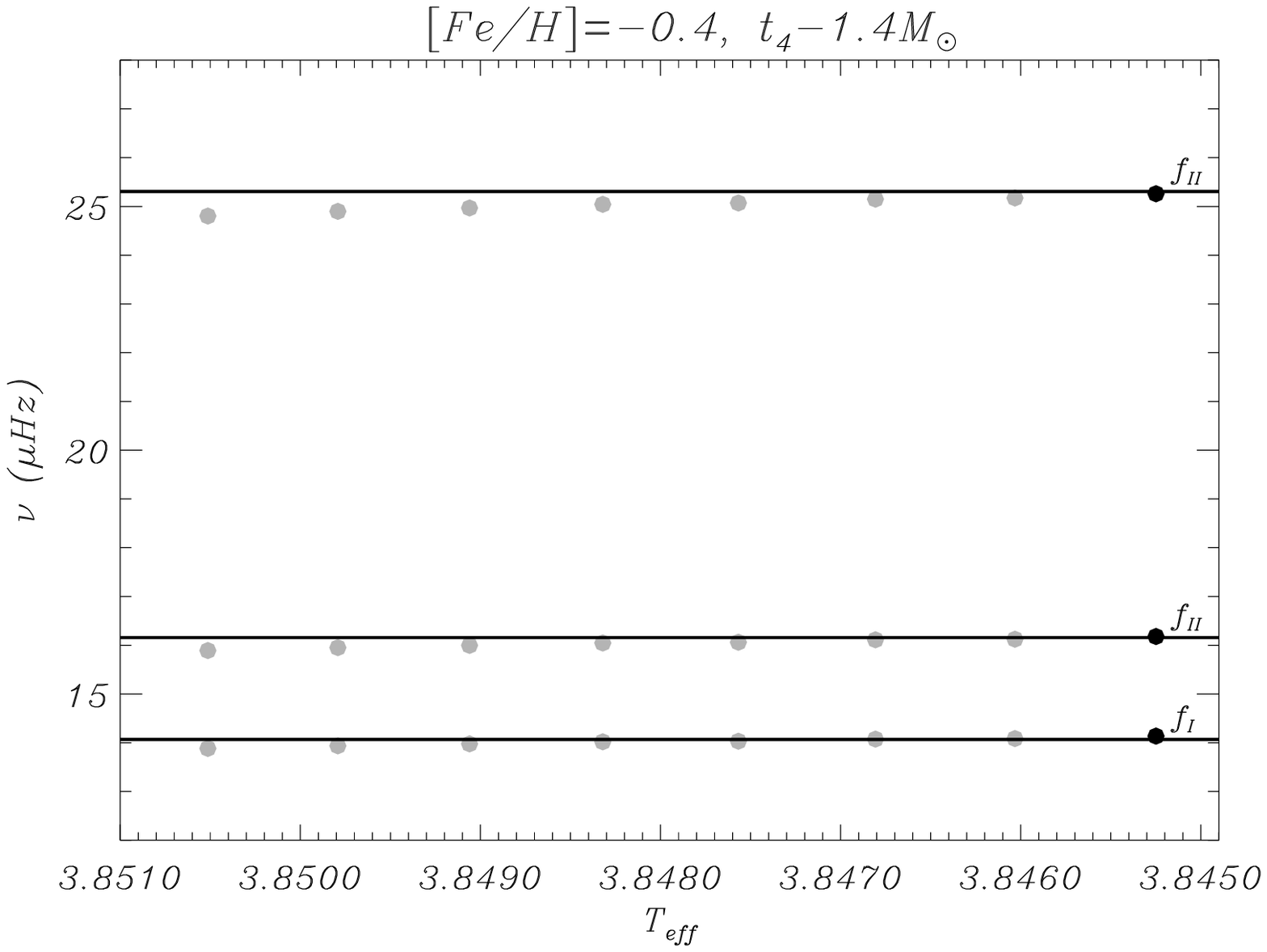}
   \includegraphics[width=9cm]{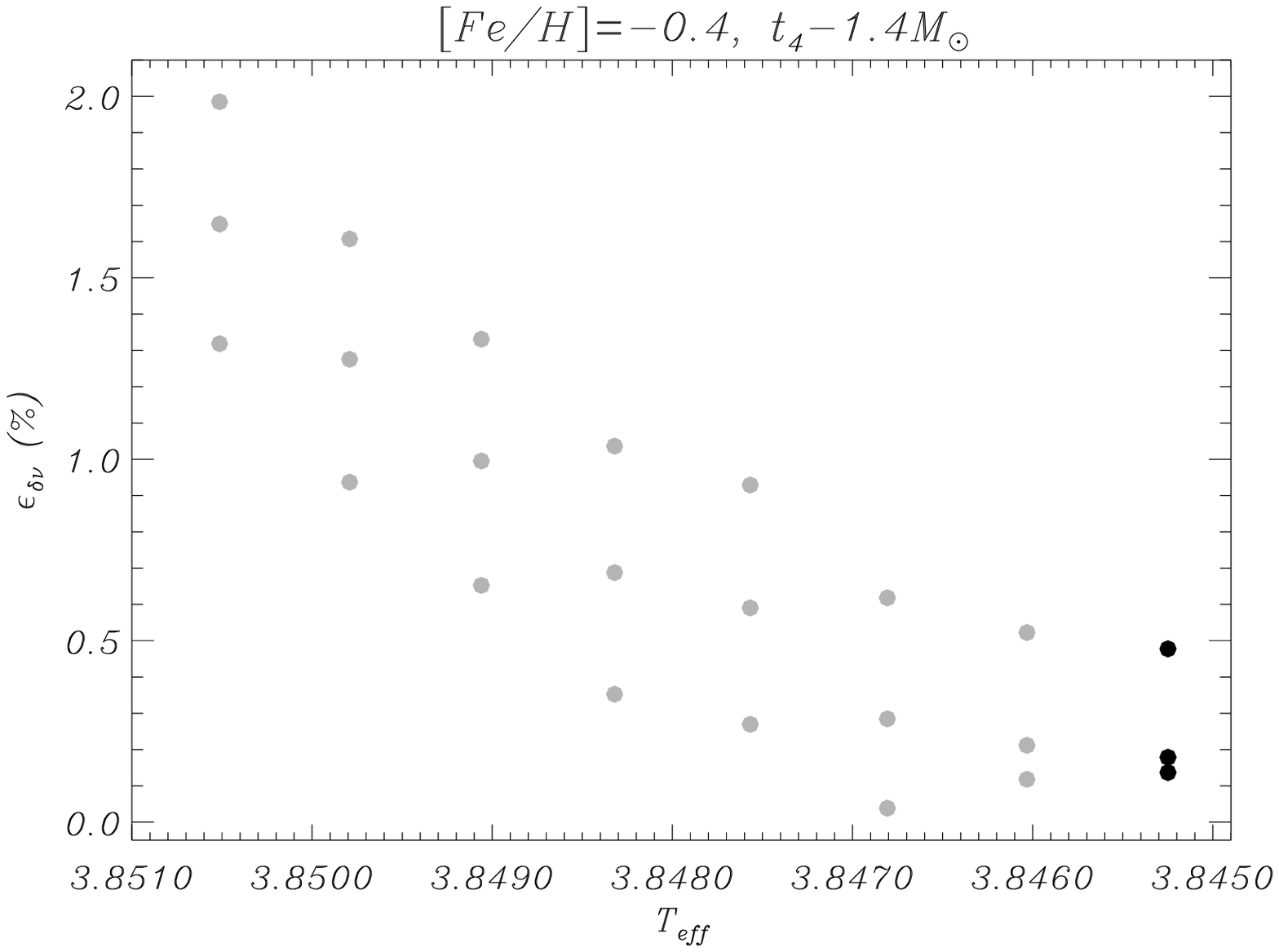}
   \caption{Evolution of theoretical frequencies (left panels)
   ($g_{26}$,$g_{41}$,$g_{47}$;$\ell=2$) corresponding to the set
   $t_4$ (see Table~\ref{tab:n-I-obs}) for selected $1.2$, $1.3$ and
   $1.4\msol$ models with a metallicity of [Fe/H]=-0.4 and ${\cal
   I}_{th}={\cal I}_{obs}(t_4)$. Filled circles represent such
   frequencies for each model within the photometric error box of
   Fig. 4. The observed frequencies $f_\mathrm{I}$, $f_{\mathrm{II}}$
   and $f_{\mathrm{III}}$ are represented by continuous lines. Right
   panels show the error $\epsilon_{\delta\nu}$ of theoretical
   frequencies with respect to the observed ones for each panel on the
   left. Frequencies of \emph{best} models are represented by black
   circles. }
  \label{fig:solu-fe04}
  \vspace{4cm}
\efige
\bfige
  \bc
   \includegraphics[width=15cm]{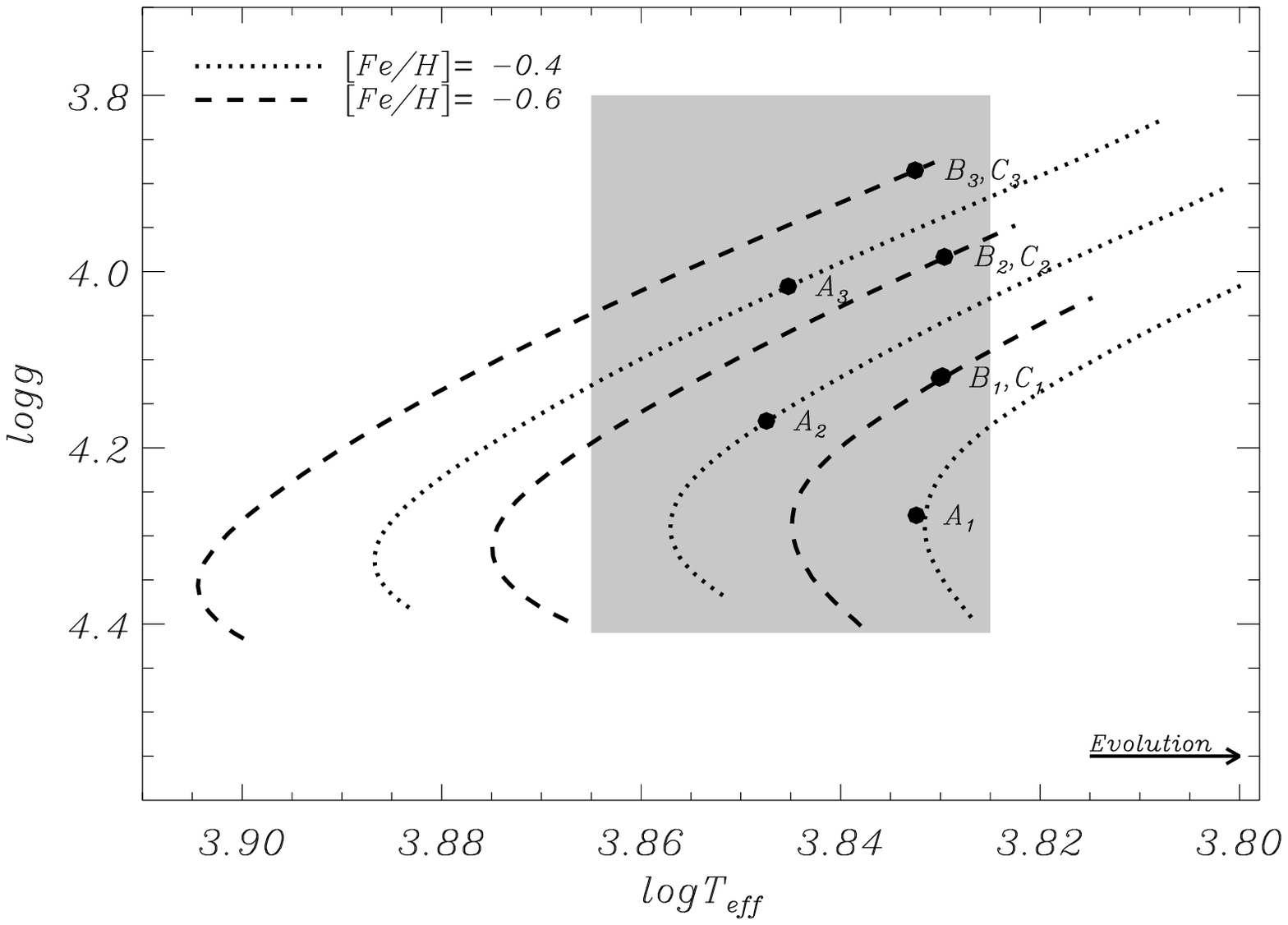}
   \caption{Surface gravity as a function of the effective temperature
            of seleceted \emph{best} models following the
            procedure given in Fig.~\ref{fig:solu-fe04}. These
            selected models are represented by filled circles which
            are labeled in Table~\ref{tab:selmodels}.  Evolutionary
            tracks of concerned models are also displayed: as in
            Fig.\ref{fig:logg-teff_general}, dotted lines represents
            tracks for models with [Fe/H]=-0.4 and dashed lines those
            for models with [Fe/H]=-0.6. The shaded area represents
            the observational error box deduced from
            Table~\ref{tab:params}. } \ec
  \label{fig:logg-teff_selmodels}
\efige
%

\begin{table*}
    \caption{Charateristics of the selected \emph{best} models obtained by 
             comparing theoretical frequencies of deduced sets from
	     representative models with ${\cal I}_{th}\approx {\cal I}_{obs}$.
	     Effective temperatures and luminosities are given on a 
	     logarithmic scale. Ages are given in Myr. For each 
	     selected model, the corresponding mode degree $\ell$ is also 
             given.}
    \vspace{1em}
    \renewcommand{\arraystretch}{1.2}
    \begin{tabular}[ht]{lccccccccc}
    \hline\hline
    & $\ell$ &  [Fe/H] & M$/\msol$ & $\teff$ & L$/\lsol$ & $\log\,g$  
    & $X_c$ & Age & ${\bar \rho}/{\bar \rho}_\odot$
 \\
    \hline
  $A_1$ & 2 & -0.4 & 1.2 & 3.83 & 0.52 & 4.28 & 0.50 & 1990 & 9.19 \\
  $A_2$ & 2 & -0.4 & 1.3 & 3.85 & 0.72 & 4.17 & 0.40 &  2100 & 6.10 \\
  $A_3$ & 2 & -0.4 & 1.4 & 3.84 & 0.90 & 4.02 & 0.26 &  2090 & 3.47 \\
  $B_1$, $C_1$ & 1,2  & -0.6 & 1.2 & 3.83 & 0.67 & 4.12 & 0.27 &  3120 & 5.36 \\
  $B_2$, $C_2$ & 1,2  & -0.6 & 1.3 & 3.83 & 0.84 & 3.98 & 0.17 &  2720 & 3.20 \\
  $B_3$, $C_3$ & 1,2 & -0.6 & 1.4 & 3.83 & 0.98 & 3.88 & 0.10 &  2290 & 2.20 \\
    \hline
    \end{tabular}
    \label{tab:selmodels}
\end{table*}

\subsection{Toward the modal identification\label{ssec:modidentif}}
    
At this stage, the choice of \emph{good} models is drastically reduced
to those within the \iobs--$\log \teff$ boxes obtained for each $t_i$
set given in Table~\ref{tab:n-I-obs}. Nevertheless it is possible
to provide further constraints on both equilibrium models and modal
identification from additional observational information.

Considering the observed metallicity for \hda\ (see
Table~\ref{tab:params}) and the corresponding error in its
determination, only models with [Fe/H]=$-0.4$ and $-0.6$ are
kept. This reduces the set of possible valid identifications to $t_1$,
$t_4$ and $t_5$ with their corresponding spherical order, $\ell$ and
\iobs\ values, shown in Table 6. This constrains the mass of models
to the range of 1.2--1.4$\msol$.

\subsubsection{Theoretical frequencies \label{sssec:freqteor}}

Another constraint for selecting the theoretical models fitting the
observations consists of comparing, for each set, the theoretical
frequencies corresponding to each $n_i$ with the observed
frequencies. To do so, theoretical oscillation spectra have been
computed in the range of $g$ modes given by selected $t_1$, $t_4$ and
$t_5$ sets for models within the error (shaded) boxes of
Fig.~\ref{fig:int-teff}.

In Fig.~\ref{fig:solu-fe04}, the set of models verifing the integral
value of $t_4$ is depicted. All these models have thus a
metallicity of $-0.4$. From top to bottom different masses are shown
($1.2$, $1.3$ and $1.4 M_{\odot}$). In left panels the frequencies for
the modes $g_{26}$, $g_{41}$ and $g_{47}$ (see chain $t_4$) obtained
with each model are represented. The horizontal lines are the observed
frequencies of \hda. From these panels it becomes apparent that there
is a model verifing fairly well the complete chain obtained by our
method as well as the observed physical characteristics. The right panels
show the relative error defined as follows:
\be
\epsilon_{\delta\nu}=\Big|\frac{\nu_i-\nu_{obs}}{\nu_{obs}}\Big|
\ee
The choice of the best model within a track can be carried out by
searching for the model minimizing this relative error. This model is
represented in Fig.~\ref{fig:solu-fe04} by filled circles.
Repeating this procedure for the rest of the selected models in
Fig.~\ref{fig:int-teff} it is possible to restrict the
complete sets of models to just nine. Seismic 
models with the same mass and metallicity but verifing modal
identifications with different $\ell$ are computed from
the same equilibrium model. This reduces the total number
of theoretical models fitting \hda\ to 6 (those verifing the complete
constraints concerning $\log \teff$, $\log g$, metallicity,
$(n,l,\sigma)$ and $\cal I_{obs}$ for each observed frequency). In
Table~\ref{tab:selmodels} the charateristics of the selected
\emph{best} models are displayed.

This final set of models is depicted in the HR diagram in Fig. 9. As
can be seen in Table~\ref{tab:selmodels}, this set is very
heterogeneous, with very different hydrogen central concentration,
age, $\log g$ and/or mean density. This is particularly useful
when additional constraints on models exist.
In this context, information on the spherical degree of modes 
(from multicolour photometry \citep{Moya04} or spectroscopy)
would be specially helpful in discriminating between models. In particular, in 
\citet{AertsCuypers04}, the three frequencies of \hda\ are suggested to
be $\ell=1$. This would restrict the representative models of the star
to $B_1$, $B_2$ and $B_3$. 

Further constraints from the pulsational behaviour of \gds\ are
likely to complete the method presented here. Particularly,
constraints on unstable modes considering a convection--pulsation
interaction can be obtained (work in preparation) from recent
theoretical developments of \citet{ahmed} and \citet{Ma04}.

%
\section{Conclusions}

By using the first order asymptotic representation for the
low-fre\-quen\-cy g-mode eigen\-values of stel\-lar pul\-sa\-tions, a
method for estimating the radial and spherical orders as well as the
\vaiss\ integral is presented. The method is developed under the
assumptions of adiabaticity and no rotation, and is based on
information gathered from ratios of observed frequencies, similary to
the large and small frequency difference for high-order $p$ modes.  It
can be shown that the method becomes useful when applied to at least 3
frequencies. For each of these observed frequencies, some
information about the $\ell$ of each mode or the assumption that all
of them have the same $\ell$ is needed.

The frequency method is specially adapted to \gds\ for two main reasons: 
1) they oscillate with $g$ modes in the asymptotic regime and 2)
their internal structure (a radiative envelope between two
convective zones, the inner core and other close to the photosphere)
make it possible to use an asymptotic expression with only one
scaling model-dependent parameter, the \vaiss\
integral. 

The self-consistency of this method is verified by means
of a theoretical exercise in which the accuracy of 
predictions given by
the expresion obtained by 
\citet{Tas80} for asymptotic $g$ modes is checked. These predictions
are compared with theoretical frequencies computed for
a model representative of a typical \gd\ star.
 

An application to the real star \hda\ is also given. This star has been
recently found to be a \gd\ star and three oscillation frequencies
have been identified \citep{AertsCuypers04}. Nine possible mode
identifications ($n$,$\ell$) as well as the corresponding \vaiss\
integral estimates are obtained.  From standard constraints on models
coming from fundamental parameters (mainly $\logg$, $\teff$ and
metallicity), nine sets of representative models are selected. From
these models, theoretical frequencies computed for the nine mode
identifications obtained are compared with observations. This lowers
the total number of representative models of \hda\ to
six. Furthermore, recent multicolor photometry results for this star
given by \citet{AertsCuypers04} suggest the spherical order to be
$\ell=1$.  This drastically reduces the number of valid models fitting
the observations to only three.

This method constitutes an important step toward the modal
identification of \gds, especially in cases when there is no
additional information, like that provided by multicolour photometry
or high resolution espectroscopy. It is, therefore, a method well
suited for the particular case of the white-light, very high precision
photometry expected to be delivered by the asteroseismology camera of
the COROT mission \citep{COROT}. Nevertheless, improvements coming from the
pulsational behaviour of \gds\, such as unstable mode predictions
considering a convection--pulsation interaction, and additional
spectroscopy or multicolor photometry information will
complete the method here presented. Therefore, this method, together 
with the additional information mentioned above, represents a complete 
scheme for studying and interpreting the observational behaviour
of $\gamma$ Doradus stars.

\begin{acknowledgements}
This work was partialy financed by the Spanish Plan Nacional del
Espacio under proyect ESP 2001-4528-PE. PJA acknowledges financial
support from the Instituto de Astrof\'{\i}sica de Andaluc\'{\i}a-CSIC by
an I3P contract (I3P-PC2001-1) funded by the European Social Fund. SM
acknowledges financial support at the same Institute from an "Averroes"
postdoctoral contract, from the Junta de Andalucia local government.
The authors would like to thank Prof. Paul Smeyers for helpful
discussions about the asymptotic theory.
\end{acknowledgements}
%



 \end{document}